# Misguided Use of Observed Covariates to Impute Missing Covariates in Conditional Prediction: A Shrinkage Problem

Charles F Manski[a] , Michael Gmeiner[b], and Anat Tambur[c]

February 2021

Abstract

Researchers regularly perform conditional prediction using imputed values of missing data. However, applications of imputation often lack a firm foundation in statistical theory. This paper originated when we were unable to find analysis substantiating claims that imputation of missing data has good frequentist properties when data are missing at random (MAR). We focused on the use of observed covariates to impute missing covariates when estimating conditional means of the form $E(y|x, w)$. Here $y$ is an outcome whose realizations are always observed, $x$ is a covariate whose realizations are always observed, and $w$ is a covariate whose realizations are sometimes unobserved. We examine the probability limit of simple imputation estimates of $E(y|x, w)$ as sample size goes to infinity. We find that these estimates are not consistent when covariate data are MAR. To the contrary, the estimates suffer from a shrinkage problem. They converge to points intermediate between the conditional mean of interest, $E(y|x, w)$, and the mean $E(y|x)$ that conditions only on $x$. We use a type of genotype imputation to illustrate.

[a]Department of Economics and Institute for Policy Research, Northwestern University, Evanston, IL 60208

[b] Department of Economics, Northwestern University, Evanston, IL 60208

[c]Feinberg School of Medicine, Northwestern University, Chicago, IL 60611

1. Introduction

Researchers regularly perform conditional prediction using imputed values of missing data. Imputations are embedded in widely used public datasets. For example, the U.S. Census Bureau provides hot-deck imputations of missing data in public releases of the Current Population Survey and other major Census surveys (U.S. Census Bureau, 2006, 2011).

Whereas the hot-deck method associates a single imputation with each case of missing data, Rubin (1987, 1996) promoted random multiple imputation (RMI) as a general approach for coping with missing values in public-use data. The adjective "random" refers to drawing imputed values at random from a specified probability distribution and treating the imputations as if they are real data. The adjective "multiple" refers to repetition of the random imputation process, generating multiple pseudo datasets and correspondingly multiple estimates of quantities of interest. Rubin (1996) made this broad recommendation (p. 473): "For the context for which it was envisioned, with database constructors and ultimate users as distinct entities, I firmly believe that multiple imputation is the method of choice for addressing problems due to missing values."

Imputation is used in medical research that aims to predict health outcomes conditional on patient covariates. Considering missing data in clinical trials, a National Research Council panel (National Research Council, 2010) cautioned against use of single imputation, but the panel argued favorably for RMI. Use of RMI has also been recommended for observational studies in medicine (e.g., Sterne *et al.*, 2009; Azur *et al.,* 2011; Pedersen *et al.*, 2017).

Various methods of genotype imputation have been proposed to increase the predictive power of medical risk assessments conditional on patient genotype (e.g., Li *et al.*, 2009). Researchers use an auxiliary database of precise genotypes for a specific subset of persons to impute precise genotypes for persons in the main study population who have only coarse genotyping.

Unfortunately, applications of imputation in general, and RMI in particular, lack a firm foundation in statistical theory. Consider RMI. Rubin originally motivated RMI from a subjective Bayesian perspective.

One places a joint subjective distribution on all observed and unobserved quantities. One wants to compute the posterior subjective distribution of some unobserved quantity conditional on all of the observed ones. Given this, Rubin's RMI is simply a computational method that uses Monte Carlo integration to approximate the mean of the posterior distribution. Appendix A explains further.

The foundational problem, that imputation conditional on observable data provides no new information beyond these data, arises when one considers RMI from a frequentist perspective, as has generally been the case in practice. Rubin asserted good frequentist properties for RMI, but he formally studied only a highly restricted form of frequentist inference; see Appendix A. Other authors have asserted broadly, but without proof, that RMI has good frequentist properties when data are missing at random (MAR). For example, Sterne *et al.* (2009) states (p. 5): "under the missing at random assumption multiple imputation should correct biases that may arise in complete cases analyses." Pedersen *et al.* (2017) states (p. 157): "Multiple imputation is implemented in most statistical software under the MAR assumption and provides unbiased and valid estimates of associations based on information from the available data."

This paper originated when we were unable to find analysis substantiating claims that RMI has good frequentist properties when data are MAR. Looking into the matter, we focused on the use of observed covariates to impute missing covariates when estimating conditional means of the form E(y|x, w). Here y is an outcome whose realizations are always observed, x is a covariate whose realizations are always observed, and w is a covariate whose realizations are sometimes unobserved.

A prominent example is hot-deck imputation in surveys, which replaces each missing value of w with an observed value of w from another respondent who has a similar value of x. Another prominent example is imputation of refined genotypes. Here w is a refined genotype that may not be observed in a study population and x is a crude type that is always observed. Then, knowledge of the distribution P(w|x) in a specific subset of persons (typically based on ethnic background) may be used to impute w in the study population.

We examine the probability limit of simple imputation estimates of E(y|x, w) as sample size goes to infinity. We find that these estimates are not consistent when covariate data are MAR. To the contrary, the

estimates suffer from a shrinkage problem. They converge to points intermediate between the conditional mean of interest, $E(y|x, w)$, and the mean $E(y|x)$ that conditions only on x. Hence, we conclude that use of observed covariates to impute missing covariates is misguided.

Section 2 presents the theoretic analysis. Section 3 uses a type of genotype imputation to illustrate.

## 2. Imputation of Missing Covariate Data in Conditional Prediction

It is well understood that imputation is not a panacea for inference with missing data. Horowitz and Manski (1998, 2000) studied nonparametric partial identification of conditional means in the absence of assumptions restricting the distribution of missing outcome and covariate data. They observed that using imputations in place of missing data does not generally yield consistent estimates, but they did not study specific imputation methods under specified assumptions on data generation. We do so here.

We examine the probability limit of estimates of conditional means that use observed covariates to randomly impute missing covariates. Formally, consider a population with members characterized by variables $(y, x, w, z)$. Here $y \in Y$ is a real outcome with bounded domain Y, whereas $x \in X$ and $w \in W$ are covariate vectors with finite domains X and W. Realizations of x are always observable, but some realizations of w are not. The binary variable z indicates whether w is observable $(z = 1)$ or not $(z = 0)$.

Let the population distribution of $(y, x, w, z)$ be denoted P. The objective is to learn $E(y|x = \xi, w = \omega)$ when $P(x = \xi, w = \omega) > 0$. A random sample of N population members are drawn. One observes $(y_i, x_i, z_i)$ for all $i = 1, \ldots, N$ and observes $w_i$ when $z_i = 1$. Suppose that, if w were always observed, one would estimate P by the empirical distribution $P_N$ and would estimate $E(y|x = \xi, w = \omega)$ by its sample analog $E_N(y|x = \xi, w = \omega)$.

Now consider estimation of $E(y|x = \xi, w = \omega)$ when some data on w are missing. We first give the general form of sample-analog estimates that treat imputed values of w as real data. We then consider estimates that impute missing values of w randomly, conditional on x.

## 2.1. Imputation in Generality

To cope with missing data on w, let each member of the population be assigned $M > 0$ imputed values $u_m \in W$, $m = 1, \ldots, M$. Single imputation occurs when $M = 1$ and multiple imputation when $M > 1$. In the sample of size N, Let $N(1, \xi, \omega)$ be the sub-sample of cases where $(z = 1, x = \xi, w = \omega)$. Let $N_m(0, \xi, \omega)$ be the sub-sample where $(z = 0, x = \xi, u_m = \omega)$. Let $N_{1\xi\omega} = |N(1, \xi, \omega)|$, $N_{m0\xi\omega} = |N_m(0, \xi, \omega)|$, and $\pi_{mN\xi\omega} \equiv N_{1\xi\omega}/(N_{1\xi\omega} + N_{m0\xi\omega})$. Then, whenever $N_{1\xi\omega} + N_{m0\xi\omega} > 0$, the $m^{th}$ imputation estimate of $E(y|x = \xi, w = \omega)$ is

$$(1) \quad \theta_{mN\xi\omega} \equiv \frac{1}{N_{1\xi\omega} + N_{m0\xi\omega}} \left( \sum_{i \in N(1, \xi, \omega)} y_i + \sum_{i \in N_m(0, \xi, \omega)} y_i \right)$$

$$= \pi_{mN\xi\omega} \frac{1}{N_{1\xi\omega}} \sum_{i \in N(1, \xi, \omega)} y_i + (1 - \pi_{mN\xi\omega}) \frac{1}{N_{m0\xi\omega}} \sum_{i \in N_m(0, \xi, \omega)} y_i .$$

Let $N \to \infty$. By the Law of Large Numbers, the probability limit of $\theta_{mN\xi\omega}$ is

$$(2) \quad \theta_{m\xi\omega} \equiv E(y|x = \xi, w = \omega, z = 1) \cdot \pi_{m\xi\omega} + E(y|x = \xi, u_m = \omega, z = 0) \cdot (1 - \pi_{m\xi\omega}),$$

where

$$(3) \quad \pi_{m\xi\omega} = \frac{P(z = 1, x = \xi, w = \omega)}{P(z = 1, x = \xi, w = \omega) + P(z = 0, x = \xi, u_m = \omega)}$$

$$= \frac{P(z = 1, w = \omega | x = \xi)}{P(z = 1, w = \omega | x = \xi) + P(z = 0, u_m = \omega | x = \xi)} .$$

In general, $\theta_{m\xi\omega}$ does not equal E(y|x, w). By the Law of Iterated Expectations,

(4)  E(y| x = ξ, w = ω)  =  E(y|x = ξ, w = ω, z = 1)·P(z = 1| x = ξ, w = ω)

  + E(y|x = ξ, w = ω, z = 0)· P(z = 0| x = ξ, w = ω).

## 2.2. Random Imputation Conditional on x

We can further characterize $\theta_{m\xi\omega}$ if we study particular classes of imputation methods under specified assumptions. We consider methods that specify a vector $G_m(u_m|x = \xi)$, $\xi \in X$ of probability distributions on W. For each person with x = ξ, the imputation method draws $u_m$ at random from $G_m(u_m|x = \xi)$.

Random imputation of genotypes as described earlier exemplifies this type of imputation. So does any deterministic imputation method that makes $u_m$ a function of x. With deterministic imputation, $G_m(u_m|x = \xi)$, $\xi \in X$ are degenerate distributions.

A first basic result holds for all specifications of $G_m$ and all processes of data generation. Whatever $G_m$ may be, $u_m$ is by construction statistically independent of (y, z) conditional on x. Hence,

(5a)  E(y|x = ξ, $u_m$ = ω, z = 0)  =  E(y|x = ξ, z = 0),

(5b)  P(z = 0, $u_m$ = ω| x = ξ)  =  P(z = 0|x = ξ)·$G_m(u_m = ω| x = ξ)$.

It follows that (2)-(3) reduce to

(6)  $\theta_{m\xi\omega}$  =  E(y|x = ξ, w = ω, z = 1)·$\pi_{m\xi\omega}$ + E(y|x = ξ, z = 0)·$(1 - \pi_{m\xi\omega})$,

(7)  $\pi_{m\xi\omega}$  =  $\dfrac{P(z = 1, w = ω| x = ξ)}{P(z = 1, w = ω| x = ξ) + P(z = 0|x = ξ)·G_m(u_m = ω| x = ξ)}$ .

This finding suggests shrinkage. Whereas E(y|x = ξ, w = ω) is a weighted average of E(y|x = ξ, w = ω,

z = 1) and E(y|x = ξ, w = ω, z = 0), θ$_{mξω}$ is a weighted average of E(y|x = ξ, w = ω, z = 1) and E(y|x = ξ, z = 0). The reason that I use the imprecise word "suggests" is that the weighting may differ in the two cases. The former weights are P(z = 1| x = ξ, w = ω) and P(z = 0| x = ξ, w = ω). The latter are π$_{mξω}$ and 1 − π$_{mξω}$.

The connection to classical shrinkage becomes exact if the missing data are MAR conditional on x, in the sense that (y, w) is statistically independent of z conditional on x. Then

(8a)    E(y|x = ξ, w = ω, z = 1)   =   E(y|x = ξ, w = ω),

(8b)    E(y|x = ξ, z = 0)   =   E(y|x = ξ),

(8c)    P(z = 1, w = ω| x = ξ)   =   P(z = 1| x = ξ)·P(w = ω| x = ξ).

It follows that (6)-(7) reduce to

(9)    θ$_{mξω}$   =   E(y|x = ξ, w = ω)·π$_{mξω}$ + E(y|x = ξ)·(1 − π$_{mξω}$),

(10)    π$_{mξω}$   =   $\dfrac{\text{P(z = 1| x = ξ)·P(w = ω| x = ξ)}}{\text{P(z = 1| x = ξ)·P(w = ω| x = ξ) + P(z = 0|x = ξ)·G}_m\text{(u}_m\text{ = ω| x = ξ)}}$ .

Thus, θ$_{mξω}$ is now a weighted average of the conditional means E(y|x = ξ, w = ω) and E(y|x = ξ). In other words, imputation shrinks estimation of E(y|x, w) toward E(y|x).

A further simplification occurs when the distribution used to impute w is G$_m$(u$_m$|x) = P(w|x), as has been the intent in genotype imputation. Then π$_{mξω}$ = P(z = 1| x = ξ) and

(11)    θ$_{mξω}$   =   E(y|x = ξ, w = ω)· P(z = 1| x = ξ) + E(y|x = ξ)·P(z = 0| x = ξ).

Hence, the asymptotic bias of the imputation estimate is [E(y|x = ξ) − E(y|x = ξ, w = ω)]·P(z = 0| x = ξ).

3. Illustration: Imputation of HLA Allele-Level Genotypes in Research Predicting Transplant Outcomes

3.1. Background

An important clinical problem in organ transplantation is to predict the outcomes that occur when an organ is transplanted into a recipient. Covariates with predictive power include data characterizing the organ and patient. A prominent consideration is the genetic match between donor and recipient, measured by their Human Leukocyte Antigen (HLA) genotypes.

A persistent problem in research predicting transplant outcomes is incomplete information on donor and recipient HLA typing. The Organ Procurement and Transplantation Network (OPTN) requires transplant centers to provide pre-transplant data to the Scientific Registry of Transplant Recipients (SRTR), which collates the data with reports of transplant outcomes and makes the combined data available for analysis. Researchers use the SRTR data to investigate how transplant outcomes vary with measured attributes of organ quality, patient age/health, and HLA typing.

Until 2010, the only HLA typing information required for donors and recipients was on the HLA (A, B, DR) loci, at the serologic level. Since then, more accurate molecular typing has been required. Over time, additional loci information has been mandated for the donor, including HLA (DQB1, DPB1, DQA1). However, no such requirements have been made for the typing of patients. Many patients are listed for transplant with only HLA (A, B, DR) low-resolution (two-digit) typing information available, although some patients have serologic equivalent DQ typing.

Higher resolution (four-digit) typing would be beneficial because each low-resolution, two-digit, typing represents multiple alleles, whereas each four-digit typing identifies a unique allele. Donor and recipient HLA types that appear matched with two-digit coding may be mismatched with four-digit coding. A patient may have antibodies to an allele within a low-resolution antigen group, but the donor typing may be of a different allele, against which the recipient does not have a donor-specific antibody. Differences at the allele level can also translate into differences in the assignment of molecular mismatches, currently proposed for use in risk stratification of transplant recipients.

Aiming to refine the predictions possible with incomplete knowledge of HLA data, methods have been developed to use available statistics on the distributions of high-resolution HLA typing within specified ethnic/national sub-populations to impute unobserved typing. Two prominent approaches are imputation of most prevalent types ("winner-take-all") and RMI. Both approaches use a known distribution of high-resolution typing given low-resolution typing. Articles imputing most prevalent types include Geneugelijk *et al.*, 2017), Tinckam *et al.* (2016), and Nilsson *et al.* (2019). Ones using RMI include Gragert *et al.* (2014) and Kamoun *et al.* (2017).

The most comprehensive database providing frequency distributions of HLA alleles and haplotypes is HaploStats ([http://www.HaploStats.org](http://www.HaploStats.org)). HaploStats facilitates access to HLA genotype frequency data for various ethnic/national groups. A researcher or clinician can input the available, incomplete, typing data for a donor or recipient who is a member of a specified group. HaploStats accesses its genotype frequency data and outputs the frequency distribution of high-resolution typing conditional on the data provided.

### 3.2. The Logic of Prediction with Incomplete HLA Data

Given that the transplant data available through the SRTR only includes low-resolution typing data, and retrospective typing of donor-recipient pairs is impractical, using HaploStats to impute high-resolution typing may seem an appealing way to overcome incompleteness of observed HLA data. However, the appeal does not survive under scrutiny.

In the notation of this paper, x is low-resolution typing available for essentially all donor-recipient pairs in the SRTR dataset, whereas w is high-resolution typing available in HaploStats. Access to HaploStats provides knowledge of the conditional distribution $P(w|x)$. Consider use of RMI. Given that w is never observed in the SRTR data, the MAR assumption necessarily holds, with $P(z = 1|x = \xi) = 0$ for all values of $\xi$. Hence, equation (11) reduces to the result $\theta_{m\xi\omega} = E(y|x = \xi)$. Thus, imputation of high-resolution

HLA does not improve prediction of transplant outcomes beyond the information in observable low-resolution SRTR data.

Imputation being ineffective, it is natural to ask what predictions of transplant outcomes can logically be made by combining SRTR and HaploStats data. Formally, what can be learned about P(y|x, w) given knowledge of P(y|x) and P(w|x)? This question has been addressed by Manski, Tambur, and Gmeiner (2019), drawing on earlier research on the *ecological inference* problem in medical risk assessment (Manski, 2018).

Analysis is simplest when the outcome of interest can take two values, say 0 and 1. For example, y = 1 may denote that a graft survives for a specified length of time and y = 0 that it does not. Then application of basic probability theory shows that, for any values (x, w) of the observed and unobserved attributes, the outcome probability P(y = 1|x = ξ, w = ω) lies between certain lower and upper bounds that are computable given the available information.

The lower bound is [P(y = 1|x = ξ) – P(w ≠ ω|x = ξ)]/P(w = ω|x = ξ) and the upper bound is P(y = 1|x = ξ)/P(w = ω|x = ξ). The lower bound is informative, in the sense of being larger than zero, if P(y = 1|x = ξ) exceeds P(w ≠ ω|x = ξ). The upper bound is informative, in the sense of being smaller than one, if P(y = 1|x = ξ) is less than P(w = ω|x = ξ).

For example, suppose the SRTR data show that when a donor and recipient have observed attributes x, the frequency with which a graft survives for a given length of time is P(y = 1|x = ξ) = 0.6. Drawing on HaploStats, suppose that w is the most prevalent pair of haplotypes when the donor and recipient have attributes ξ, with P(w = ω|x = ξ) = 0.8. Then the computable lower bound on P(y = 1|x = ξ, w = ω) is (0.6 – 0.2)/0.8 = 0.5 and the upper bound is 0.6/0.8 = 0.75.

Going beyond the case where the outcome is binary, the ecological inference problem has been studied when the objective is to learn the conditional mean E(y|x, w) for a real-valued outcome. The analysis is mathematically more subtle than when y is binary, but a tractable finding emerges. Knowledge of P(y|x) and P(w|x) yields a computable bound on E(y|x, w). See Manski (2018).

3.3. Using Imputation in Risk-Assessment Models that Condition on Number of HLA Mismatches

Analysis of the ecological inference problem proves that imputation of HLA types cannot be informative about the distribution P(y|x, w) of transplant outcomes when x is observed typing, w is unobserved typing, and knowledge of P(w|x) is used to impute w. This fact does not, however, imply that imputation is useless for all versions of transplant risk assessment.

Transplant researchers often aim to learn not P(y|x, w) but rather P[y|f(x, w)], where f(·, ·) is a specified many-to-one function of (x, w). In particular, it is common to predict graft survival conditional on the number of (donor, recipient) HLA mismatches at various loci, rather than on the underlying HLA types. Many combinations of types can yield the same number of mismatches.

When researchers perform risk assessment with models that condition only on the number of mismatches, it is theoretically possible that imputations may have predictive power. The frequency distributions of high-resolution types generated by HaploStats condition on all of the low-resolution HLA data that researchers input, not only on the number of mismatches implied by these data. Hence, imputations in principle might add predictive power to that attainable conditioning only on number of mismatches.

It does not seem possible to determine theoretically whether imputations have predictive power when used in risk-assessment models that condition only on numbers of mismatches rather than on the totality of observed HLA data. We can, however, use available SRTR data to illustrate such use of imputation. We summarize here and provide details in Appendix B.

We consider use of a logit model to predict five-year graft survival as a function of the numbers of low-resolution mismatches at the (A, B, DR) loci. This model is estimable with the SRTR data. The estimate presented in the appendix shows that the probability of five-year graft survival decreases with the number of mismatches at each of the three loci, with DR mismatch having the strongest and statistically most significant effect.

Now suppose that one were to use randomly imputed values of DR rather than actual DR data when estimating the logit model. As described in Appendix B, we use the SRTR data, which contains actual low-resolution (A, B, DR) data, to determine the empirical conditional distribution P(DR|A, B). This done, we perform RMI, repeatedly imputing DR values and estimating the logit model. We find that imputation of DR does not reveal the actual strong effect of DR mismatch on graft survival. To the contrary, the mean of the RMI coefficients for DR mismatch is close to zero. Thus, imputation is not informative in this illustration of risk assessment conditional on numbers of mismatches.

## Appendix A. Rubin's Bayesian and Frequentist Theory of RMI

A concise statement of the Bayesian theory motivating RMI was given in Rubin (1996), where he considered the posterior distribution for a real parameter Q(Y), Y being a random vector with some components observed and some missing. He wrote (p. 476).

"The key Bayesian motivation for multiple imputation is given by result 3.1 in Rubin (1987). . . . . the results and its consequences can be easily stated using the simplified notation that the complete-data are $Y = (Y_{obs}, Y_{mis})$, where $Y_{obs}$ is observed and $Y_{mis}$ is missing. Specifically, the basic result is

$$P(Q|Y_{obs}) = \int P(Q|Y_{obs}, Y_{mis})P(Y_{mis}|Y_{obs})dPY_{mis}."$$

In Bayesian language, $P(Q|Y_{obs})$ is the posterior predictive distribution of Q conditional on $Y_{obs}$, $P(Q|Y_{obs}, Y_{mis})$ is the posterior for Q given $(Y_{obs}, Y_{mis})$, and $P(Y_{mis}|Y_{obs})$ is the posterior for $Y_{mis}$ given $Y_{obs}$. In non-Bayesian language, the equation applies the Law of Total Probability. Rubin supposed that $P(Q|Y_{obs}, Y_{mis})$ and $P(Y_{mis}|Y_{obs})$ are specified subjective distributions, making $P(Q|Y_{obs})$ computable. In practice, he focused on the posterior mean of Q; that is $E(Q|Y_{obs}) = \int E(Q|Y_{obs}, Y_{mis})P(Y_{mis}|Y_{obs})dPY_{mis}$.

Observe that Rubin's "basic result" does not explicitly refer to RMI. He interpreted it as RMI by considering Monte Carlo integration as a practical approach to approximate $E(Q|Y_{obs})$. To perform Monte Carlo integration, one draws repeated values of $Y_{mis}$ at random from $P(Y_{mis}|Y_{obs})$ and averages the resulting values of $E(Q|Y_{obs}, Y_{mis})$. Semantically, one may refer to Monte Carlo draws of $Y_{mis}$ as imputations. Hence, RMI is Monte Carlo integration.

The above motivation for RMI is well-grounded from a subjective Bayesian perspective. A disconnect between the theory and practice of RMI stems from the effort made by Rubin to assert desirable frequentist properties for RMI. To a subjective Bayesian, the posterior mean $E(Q|Y_{obs})$ is well-defined and interpretable regardless of whether it equals an objective quantity of scientific interest. A frequentist, however, assumes the existence of an objective quantity of interest, say $Q^*$, and wants to estimate this quantity well in some sense, across repeated samples.

In general, the posterior mean $E(Q|Y_{obs})$ need not be a good estimate of $Q^*$ when $P(Q|Y_{obs}, Y_{mis})$ and $P(Y_{mis}|Y_{obs})$ are simply subjective distributions. To prove good frequentist properties for $E(Q|Y_{obs})$ typically requires one to assume that $P(Q|Y_{obs}, Y_{mis})$ and $P(Y_{mis}|Y_{obs})$ are objectively correct. Rubin demonstrated awareness of this core requirement when he wrote (Rubin, 1996, p. 474): "My conclusion is that 'correctly' modeling the missing data must be, in general, the data constructor's responsibility." However, he provided no evidence that data constructors are able to model missing data correctly.

Rubin argued that two desirable frequentist properties for statistical procedures are "randomization validity," which he interpreted as requiring approximately unbiased point estimates of scientific estimands, and "confidence validity," requiring that actual coverage probabilities for confidence intervals should be at least as large as nominal coverage probabilities. He wrote (Rubin, 1996, p. 476):

"Multiple imputation was designed to satisfy both achievable objectives by using the Bayesian and frequentist paradigms in complementary ways: the Bayesian model based approach to *create* procedures, and the frequentist (randomization-based approach) to *evaluate* procedures."

Continuing, he wrote that if the multiple imputations are "proper" and complete data inference is randomization-valid, then (p. 477): "the large-*m* repeated-imputation inference . . . is randomization-valid for the scientific estimand *Q, no matter how complex the survey design.*"

It is not easy to understand Rubin's extended verbal discussion of what he means by "proper" multiple imputation. However, we believe that we understand the type of frequentist inference that he had in mind. His symbol *m* refers to the number of random draws made from $P(Y_{mis}|Y_{obs})$ and, hence "large-*m*" refers to asymptotic analysis as *m* goes to infinity. Thus, he meant that, by the Law of Large Numbers and the Central

Limit Theorem, Monte Carlo integration yields a well-behaved estimate of a population mean as the number of pseudo-draws goes to infinity. Randomization validity in this sense means that RMI yields a consistent estimate of $E(Q|Y_{obs})$ asymptotically in m. It implies nothing about the quality of RMI in estimation of $Q^*$.

## Appendix B. Empirical Illustration of RMI Applied to HLA Genotypes Types

As in Manski, Tambur, and Gmeiner (2019), we examine data on the outcomes of deceased-donor transplants recorded in the SRTR from 2009 through 2018. Most HLA codings in the SRTR are at the two-digit level. We convert occasional four-digit codings to two-digit following OPTN guidelines with the exception of the coding 103 for DR. Transplants with codings 2, 3, 5, or 6 for DR are excluded due to ambiguity of these codings. Mismatches are defined as the number of unique antigens the donor has that the recipient does not have.

We study transplants for which the donor and patient were both coded as white. Within this population, we consider each combination of (A, B, DR) antigens separately for the donor and recipient. Rather than use HaploStats to impute two-digit DR genotypes, we use the available SRTR data on (A, B, DR) types. We generate separate estimates of P(DR|A, B) for donors and recipients

Whereas some (A, B) types are common among SRTR transplants with white donors and recipients, other types are sparse. To obtain meaningful estimates of P(DR|A, B), we restrict attention to (A,B) types for which at least 10 observations were present in the data. For each such case, we compute the empirical distribution of DR conditional on (A, B) and use it as the estimate of P(DR|A, B).

When analyzing transplant outcomes, we restrict the sample to adult transplants in which the patient was receiving their first transplant and the KDPI variable can be calculated. We define five-year survival to take the value 0 if an individual has re-transplant, death, or otherwise graft failure within 1,825 days of transplant. The five-year survival variable takes the value 1 if the individual is observed for more than 1,825 days after transplant and the first date of a failure event, if any, occurs after 1,825 days. Our final estimation sample, comprising cases for which five-year survival can be calculated, and for which the A-B combination of both patients and donors has more than 10 observations, contains 5,045 transplants.

Coefficients from logit regressions using observable (A, B, DR) data are in table 1. We find that both B and DR mismatch have negative association with survival probability, the coefficients being strong in magnitude and statistically significant by conventional criteria.

Table 1: Logit Coefficients Using Observed Data

|  | Five Year Survival |
| --- | --- |
| A Mismatches [0-2] | 0.036 |
|  | (0.051) |
| B Mismatches [0-2] | -0.118 |
|  | (0.051) |
| DR Mismatches [0-2] | -0.163 |
|  | (0.049) |
| Constant | 0.974 |
|  | (0.053) |
| N | 5,045 |

Robust standard errors in parentheses

We next replace observable DR data with imputations. We perform RMI with 50 repetitions, imputing DR separately for donors and patients using the empirical distributions, P(DR|A,B). In each repetition, we estimate the logit model in Table 1, again using observed mismatches for A and B, but now using imputed DR mismatches. The means and standard deviations of the logit coefficients across the 50 repetitions are shown in table 2.

Table 2: Average Logit Coefficients using RMI to Impute DR Mismatches

| A Mismatches | .008 |
| --- | --- |
|  | (.003) |
| B Mismatches | -.194 |
|  | (.010) |
| DR Mismatches | .003 |
|  | (.047) |

Averages are from 50 RMI draws. Standard deviations in parentheses.

The primary insight is that the average coefficient for DR mismatches is near 0 when using imputed data rather than actual data. Also note that the average coefficients for A and B mismatches are slightly more negative than the analogous coefficients in Table 1. This suggests that the correlation between DR and other mismatches causes the effect of DR to load onto the other mismatch coefficients when the true DR data are not utilized.